\documentclass[reprint, aip, jap, amsmath, amssymb]{revtex4-1}

\usepackage{amsmath}
\usepackage{graphicx}
\usepackage[separate-uncertainty=true]{siunitx}
\usepackage[hidelinks]{hyperref}

\usepackage[utf8]{inputenc}
\usepackage[T1]{fontenc}
\usepackage{mathptmx}

\renewcommand{\d}[1]{\mathrm{d}#1}
\newcommand{\dd}[2]{\frac{\mathrm{d}#1}{\mathrm{d}#2}}
\newcommand{\pdd}[2]{\frac{\partial #1}{\partial #2}}

\newcommand{\ac}{a_\mathrm{c}}
\newcommand{\ad}{a_\mathrm{d}}
\newcommand{\Cl}{C_\mathrm{l}}
\newcommand{\Cs}{C_\mathrm{s}}
\newcommand{\Csat}{C_\mathrm{sat}}
\newcommand{\Dg}{D_\mathrm{g}}
\newcommand{\Ds}{D_\mathrm{s}}
\newcommand{\ddiff}{d_\mathrm{diff}}
\newcommand{\fc}{f_\mathrm{c}}
\newcommand{\fp}{f_\mathrm{p}}
\newcommand{\gammaLG}{\gamma_\mathrm{LG}}
\newcommand{\gammaSG}{\gamma_\mathrm{SG}}
\newcommand{\gammaSL}{\gamma_\mathrm{SL}}
\newcommand{\hint}{h_\mathrm{int}}
\newcommand{\nd}{n_\mathrm{d}}
\newcommand{\ndiff}{n_\mathrm{diff}}
\newcommand{\nedge}{n_\mathrm{edge}}
\newcommand{\nsat}{n_\mathrm{sat}}
\newcommand{\ratio}{{\tauevap/\taudiff}}
\newcommand{\tauevap}{\tau_\mathrm{evap}}
\newcommand{\taudiff}{\tau_\mathrm{diff}}
\newcommand{\td}{t_\mathrm{d}}
\newcommand{\tevap}{t_\mathrm{evap}}
\newcommand{\thetaeq}{\theta_\mathrm{eq}}
\newcommand{\thetar}{\theta_\mathrm{r}}
\newcommand{\Vd}{V_\mathrm{d}}
\newcommand{\Vdepin}{V_\mathrm{depin}}

\graphicspath{{Figures/}}
\begin{document}
\newlength{\figwidth}
\setlength{\figwidth}{0.6\linewidth}
\setlength{\figwidth}{\linewidth}

\title{Compound redistribution due to droplet evaporation\\on a thin polymeric film: theory}

\author{Thijs W.G. van der Heijden}
\email{t.w.g.van.der.heijden@tue.nl}
\affiliation{Department of Applied Physics, Eindhoven University of Technology, P.O. Box 513, 5600 MB Eindhoven, The Netherlands}

\author{Anton A. Darhuber}
\affiliation{Department of Applied Physics, Eindhoven University of Technology, P.O. Box 513, 5600 MB Eindhoven, The Netherlands}

\author{Paul van der Schoot}
\affiliation{Department of Applied Physics, Eindhoven University of Technology, P.O. Box 513, 5600 MB Eindhoven, The Netherlands}
\affiliation{Instituut voor Theoretische Fysica, Universiteit Utrecht, Princetonplein 5, 3584 CC Utrecht, The Netherlands}

\date{\today}

\begin{abstract}
\noindent 
A thin polymeric film in contact with a fluid body may leach low-molecular-weight compounds into the fluid. If this fluid is a small droplet, the compound concentration within the liquid increases due to ongoing leaching in combination with the evaporation of the droplet. This may eventually lead to an inversion of the transport process and a redistribution of the compounds within the thin film. In order to gain an understanding of the compound redistribution, we apply a macroscopic model for the evaporation of a droplet and combine that with a diffusion model for the compound transport. In the model, material deposition and the resulting contact line pinning are associated with the precipitation of a fraction of the dissolved material. We find three power law regimes for the size of the deposit area as a function of the initial droplet size, dictated by the competition between evaporation, diffusion and the initial compound concentrations in the droplet and the thin film. The strength of the contact line pinning determines the deposition profile of the precipitate, characterised by a pronounced edge and a linearly decaying profile towards the centre of the stain. Our predictions for the concentration profile within the solid substrate resemble patterns found experimentally.
\end{abstract}

\maketitle

The following article will be submitted to Journal of Applied Physics. After it is published, it will be found at \mbox{\url{https://aip.scitation.org/journal/jap}}.

\section{Introduction}
The process of staining of surfaces due to the evaporation of fluid droplets containing dissolved or suspended material finds its most prominent example in the so-called coffee stain effect~\cite{Deegan1997, Deegan2000}. The formation of the dark rings of deposit is a result of the material being transported by internal flows inside the droplet caused by differences in the evaporation rate across the surface. Apart from a fundamental scientific interest, understanding the deposition of solids and being able to predict the topology of the resulting stain is of importance in applications such as inkjet printing~\cite{Derby2010, He2017}, semiconductor device manufacturing~\cite{Wei2009, Belmiloud2012, Belmiloud2014} and the creation of colloidal photonic crystals~\cite{Norris2004}. The redistribution and deposition of material due to the evaporation of a droplet are therefore studied extensively, both experimentally~\cite{Deegan1997, Deegan2000, Marin2011, Eral2011, Berteloot2012, Eral2013} and theoretically~\cite{Deegan1997, Deegan2000, Berteloot2012, Kang2016, Man2016, Xie2018, Kolegov2019}. 

Much less studied, although also of significant industrial interest, is the redistribution of material that originates from the substrate onto which a droplet has been deposited. For instance, in the semiconductor industry, 
water droplets left behind during immersion photolithography may disrupt the designed structures by redistributing compounds in the photosensitive polymer layer~\cite{Wei2009,Chang2006}. Arguably, these so-called watermark defects are caused by compounds dissolving in the drop. During the course of the evaporation of the droplet, these compounds may diffuse back into the substrate or precipitate onto the surface. 

In this manuscript, we investigate this phenomenon theoretically. We find that the size of the deposit stain depends on the initial droplet size, where small droplets produce a deposit that in comparison is larger than those left behind by large droplets. We relate this to the relative timescales of evaporation and diffusion, and whether or not the fluid itself contains contaminants. If deposition of material occurs only near the contact line of the shrinking droplet, which may as a result become pinned, our macroscopic model predicts that the deposit profile decays linearly from the edge towards the centre. The height of the rim is determined by the strength of the pinning. Whilst pinning and depinning in our theory are symmetric by construction, this needs not be the case in practice. Depinning and the final stages of drying frequently break circular symmetry~\cite{Pittoni2013, Belmiloud2014}. In spite of its simplicity, our model predictions do qualitatively describe the salient features of the structure of watermark defects seen experimentally~\cite{Chang2006, Wei2009}.

The remainder of this paper is organised as follows. In Section~\ref{sec:theory}, we present our model for compound redistribution, in which we combine a macroscopic description for the evaporation of a droplet and diffusive transport of compounds between the thin film and the fluid. Section~\ref{sec:deposit} describes our findings relating the deposit to the initial droplet size. Because in our model contact line pinning is directly linked to precipitation of dissolved compounds, we find that the solubility limit is a crucial quantity in relation to the size and topology of the deposit stain. Section~\ref{sec:watermarks} presents an overview of the redistribution of material within the thin film and relates this to observations on watermark defects. In Section~\ref{sec:conclusions}, we summarise our results and present our main conclusions.

\section{Theory}
\label{sec:theory}
For simplicity, we consider droplets that are smaller than the capillary length $l_\mathrm{c} = \sqrt{\gammaLG/\rho g}$, where $\gammaLG$ denotes the liquid-gas interfacial tension, $\rho$ the mass density of the liquid and $g$ the gravitational acceleration. Water in air at room temperature has a capillary length of $l_\mathrm{c}\simeq \SI{3}{\milli\metre}$, implying that droplets that are smaller than this must have an equilibrium shape of a spherical cap~\cite{DeGennes2004}. If a droplet is deposited onto a thin solid film, compounds from that thin film could diffuse into the liquid. The liquid itself need not be pure either and may have dissolved compounds in it. For definiteness, we presume these compounds to be of the same species as those that diffuse out of the substrate into the droplet. This is not really a restriction, as the diffusion processes for multiple compounds occur independently of each other, if their concentrations are sufficiently low, in which case the transport processes must be additive. Our model compound has a solubility limit, implying that it falls out of solution if the concentration exceeds that limit.

In order to describe the droplet dynamics during evaporation, we may use a simple macroscopic model, such as presented in \citet{VanderHeijden2018}. This model combines the relaxation of the droplet shape, diffusive evaporation and contact line pinning of the droplet, and describes the observed dynamics of evaporating droplets remarkably well. For simplicity, however, we presume here that a droplet is deposited on the substrate at its equilibrium angle, and consequently the droplet shape relaxation does not directly play a role in our description. In that case, our model needs only three ingredients: (1) the diffusive evaporation at the level of the steady-state diffusion of the fluid into the ambient atmosphere~\cite{Picknett1977,Erbil2002}, (2) diffusive exchange of compounds between the droplet and the substrate, and (3) a contact line pinning mechanism described by the imbalance between the instantaneous capillary forces and a pinning force, presumed to be spatially homogeneous~\cite{DeGennes1985,Stauber2015a,Snoeijer2013}. In our model, the pinning force is zero until precipitation occurs, that is, when the compound concentration in the fluid drop reaches the solubility limit. The magnitude of the pinning force we take as a model parameter, independent of the amount and distribution of the precipitate on the substrate.

After the droplet is deposited on the solid substrate, compounds present in the substrate may diffuse into the liquid or \textit{vice versa}. We take vertical diffusion to be the dominant transport mechanism. This is plausible in the limit of sufficiently thin solid films, that is, films that are much thinner than the characteristic size of the deposited droplet. Indeed, if the  radial and vertical diffusivities are equal, then the ratio of the relevant timescales related to (1) the diffusion of material from the film into the droplet or \textit{vice versa} and (2) the radial equilibration within the film, scales with the square of the ratio of the pertinent length scales. For the film this is the film thickness, whilst that for the droplet is the initial droplet radius. Typical film thicknesses in semiconductor photolithography are in the order of \SI{100}{\nano\metre} and initial droplet sizes are in the range of tens to hundreds of micrometres~\cite{Chang2006, Kusumoto2006, Streefkerk2006}, so the timescales differ by four to six orders of magnitude, and our assumption is reasonable.

In the late stages of the evaporation of the droplet, the concentration of dissolved material within the drop increases due to the decrease of the droplet volume. In reality, because of this, the diffusive transport inverts the material flow, causing material transport back into the film. In addition, the concentration may reach the solubility limit $\Csat$ of the compound in the fluid, leading to precipitation. From this point on, the compound concentration in the droplet remains constant at the saturation level. Due to internal flows in the droplet, the precipitate arguably accumulates near the contact line~\cite{Deegan1997, Man2016}. This implies that for a pinned contact line the stain becomes ring-like. We find that if the contact line pinning can be overcome by the capillary forces acting on the contact line, the contact line moves inward during evaporation. The height of the deposit then decreases linearly towards the centre of the stain. The height of the ring around the edge of the smeared-out stain depends on the strength of the contact line pinning. Experiments on the topology of the stains support our deposition model that in fact is inspired by the watermark creation mechanism proposed by~\citet{Belmiloud2012}.

\begin{figure}[htp!]
\centering
\includegraphics[width=\figwidth]{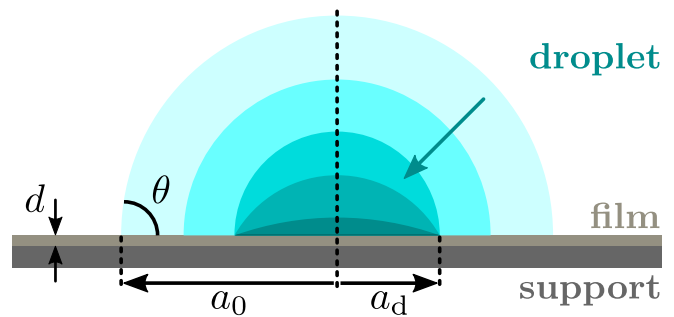}
\caption{Sketch of the droplet with initial size $a_0$ and contact angle $\theta$ evaporating on the thin film with thickness $d$. The size $\ad$ of the deposition area is indicated. The thin film rests on an impermeable support.}
\label{fig:sketch_droplet}
\end{figure}

If a droplet is deposited on a surface and is left to evaporate, the radius $a$ of the contact area decreases as the volume of fluid decreases. We envisage isothermal, quasi steady-state diffusive evaporation into an infinite ambient atmosphere, and let the contact angle $\theta$ remain constant during the evaporation until precipitation occurs, see Fig.~\ref{fig:sketch_droplet}. 
In that case, the time dependence of the contact area radius $a$ follows the familiar square root law~\cite{Picknett1977,Erbil2002,Popov2005}, 
\begin{equation}
\label{eq:a_t}
a(t) = \sqrt{a_0^2 - \Lambda(\theta) t},
\end{equation}
where $a_0=a(0)$ denotes the initial contact radius, and 
\begin{equation}
\label{eq:Lambda}
\Lambda(\theta) = \frac{4 \Dg \Delta c}{\rho} \frac{f(\theta) \sin^2\theta}{2-3\cos\theta+\cos^3\theta},
\end{equation}
is a measure for the evaporation rate. Here, $\Dg$ denotes the vapour diffusion coefficient, and $\Delta c \equiv c_\mathrm{s} - c_\infty$ is the difference between the vapour concentration close to the surface $c_\mathrm{s}$ (in units of mass per volume), which as usual we presume to be at the saturation level, and the vapour concentration of the ambient atmosphere $c_\infty$. $\rho$ denotes as before the mass density of the liquid and $f(\theta)$ is a geometric factor that relates the droplet shape to the evaporation rate. We do not reproduce the lengthy expression for $f(\theta)$ here, but instead refer the reader to Refs.~\cite{Picknett1977, Erbil2002, Popov2005}. From Eq.~\eqref{eq:a_t} we deduce that there must be a characteristic timescale $\tauevap$ for the evaporation process, which we take to be the maximum time it takes for a droplet to fully evaporate. For a droplet with a given initial volume $V_0$, this happens to be the case for a contact angle $\theta=\pi/2$, for which we find,
\begin{equation}
\label{eq:tau_evap}
\tauevap = \frac{a_0^2}{\Lambda(\pi/2)} = \frac{\rho a_0^2}{2\Dg \Delta c} = \frac{\rho}{2 \Dg \Delta c}\left(\frac{3V_0}{2\pi}\right)^{2/3}.
\end{equation} 

As the compounds dissolved in the liquid precipitate, the contact line may become pinned onto the precipitate with some pinning force \textit{per unit length} $\fp$. $\fp$ we treat as a model parameter that does not depend on any spatial position and hence is not connected with irregularities on the precipitate. For the duration of the pinning, the contact area radius $a$ remains constant in time, causing the contact angle $\theta$ to decrease. As the droplet further evaporates, the capillary force \textit{per unit length} $\fc$ obeys
\begin{equation}
\label{eq:fc}
\fc = -\gammaLG\left(\cos\theta-\cos\thetaeq\right),
\end{equation}
showing that the capillary force increases in magnitude as $\theta$ decreases~\cite{DeGennes1985, Bonn2009, Stauber2015a}. In Eq.~\eqref{eq:fc}, $\thetaeq$ denotes the thermodynamic equilibrium value of the contact angle $\theta$, described by Young's equation,
\begin{equation}
\label{eq:young}
\gammaSG - \gammaSL - \gammaLG\cos\thetaeq = 0,
\end{equation}
where $\gammaSG$ and $\gammaSL$ denote the solid-gas and solid-liquid interfacial tensions, respectively. At the moment the capillary force $\fc$ exceeds the pinning force $\fp$, the droplet depins and the contact line proceeds to move inwards again. The radius again obeys Eq.~\eqref{eq:a_t}, albeit with a different contact angle $\thetar$, and $a_0$ now refers to the radius at the point in time of depinning that then defines where we reset time to naught. We refer to $\thetar$ as the \textit{receding} contact angle, defined as the angle at which the pinning and capillary forces balance,
\begin{equation}
\label{eq:theta_r}
\thetar = \arccos \left(\cos\thetaeq + \fp/\gammaLG\right).
\end{equation}
Note that the droplet becomes permanently pinned if $\cos\thetaeq + \fp/\gammaLG > 1$.

As already alluded to, during the evaporation process, compounds present in the thin film may diffuse into the liquid droplet and compounds in the liquid droplet may diffuse into the film. As we presume vertical diffusion to be the dominant transport mechanism, the time evolution of the concentration profile $\Cs(r,z,t)$ within the thin film obeys the one-dimensional diffusion equation,
\begin{equation}
\label{eq:diff}
\pdd{\Cs}{t}=\Ds \pdd{^2\Cs}{z^2},
\end{equation} 
where $r$ denotes the radial position from the centre of the liquid droplet, $z$ is the vertical position in the film and $\Ds$ is the diffusion coefficient of the compound in the thin film. In our model description, $\Ds$ depends on neither the radial nor the vertical position. In reality, diffusion may be anisotropic and depend on, \textit{e.g.}, the water content inside the substrate, the local temperature or the internal structure of the film. Notice that although there is no radial component in the diffusive transport, the concentration in the film becomes a function of $r$ because the droplet deposited onto the film evaporates and hence becomes smaller with time.

The characteristic time scale $\taudiff$ associated with the diffusion of compounds within the thin film may be estimated by the time it takes for the compounds to traverse the thickness $d$ of the film,
\begin{equation}
\label{eq:tau_diff}
\taudiff = \frac{d^2}{2 \Ds}.
\end{equation}
We calculate the amount of material that diffuses from the thin film into the liquid (or \textit{vice versa}) using the flux density $J_z(r,t)$ along the normal of the interface from
\begin{equation}
\label{eq:flux_density}
J_z (r,t) = -\Ds \left.\pdd{\Cs}{z}\right|_{z=d}.
\end{equation}
This expression ignores any differences in the affinity of the compounds for film and liquid, but has the advantage that it is the simplest description of the diffusive process. The flux into the support of the thin film we put equal to zero, hence
\begin{equation}
\left.\pdd{\Cs}{z}\right|_{z=0} = 0.
\end{equation}

Due to the transport of material between the thin film and the liquid droplet, as well as the ongoing evaporation of the droplet, the concentration of the compounds within the droplet changes as a function of time and potentially of position. However, as the evaporation causes internal flows within the droplet~\cite{Hu2005, Ristenpart2007, Girard2008}, we presume that mixing is instantaneous on the timescale of the diffusive processes involving the thin film. This then implies that the compound concentration $\Cl=\Cl(t)$ within the liquid is independent of the position in the drop. This is a reasonable approximation, provided that the droplet shape relaxation time is much shorter than both the evaporation time and the diffusive timescale of the compound in the thin film~\cite{VanderHeijden2018}. 

The challenge now is to link the concentration of compounds in the droplet $\Cl$ to a boundary condition for the diffusion equation of those compounds in the thin film. Because we presume the concentration of compounds in the droplet to be homogeneous, we cannot impose a Neumann boundary condition, which would be the most natural boundary condition in the context of mass transport. This forces upon us a (time-dependent) Dirichlet boundary condition,
\begin{equation}
\lim_{z\to d}\Cs(r,z,t)=\Cl(t).
\end{equation}
This implies that at $t=0$ an infinitesimally narrow region in the film has the same compound concentration as the fluid. Consequently, for short times we may well be overestimating the material transport between thin film and fluid. This we do not see as a serious drawback, for it is consistent with our initial condition, which is a step function,
\begin{equation}
\Cs(r,0\leq z<d,0)=C_0(r),
\end{equation}
with $C_0(r)$ the initial concentration of the compounds in the film. In the case of an initially patterned substrate, such as is the case, \textit{e.g.}, after irradiation of a photoresist in photolithography~\cite{Wei2009}, $C_0$ depends on the position $r$. For simplicity, we presume any patterning to be radially symmetric. As to be discussed in more detail below, any unphysical initial response relaxes relatively fast on the timescale of the diffusive processes that we focus on.

We need to numerically solve for the two quantities that describe the compound concentration in the thin film and in the drop, $\Cs(r,z,t)$ and $\Cl(t)$. The latter we calculate at every time step, by taking the ratio of the amount of dissolved material and the instantaneous droplet volume $V(t)$,
\begin{equation}
\Cl(t) = \frac{\Cl(0)V(0) + \int_0^t \d{t'} \int_0^{a(t')} \d{r}\, 2 \pi r J_z(r,t')}{V(t)},
\end{equation}
provided that the concentration is smaller than the saturation value, $\Cl(t)<\Csat$. Once the concentration has reached the saturation value, it remains constant, $\Cl(t)=\Csat$, for all further times. The surplus material is deposited onto the thin film near the contact line of the drop. We keep track of the amount of deposited material as a function of time and radial position.

We solve Eq.~\eqref{eq:diff} by adding and subtracting from $\Cs(r,z,t)$ the instantaneous droplet concentration $\Cl(t)$, and define an auxiliary function $\Cs(r,z,t)-\Cl(t)$ and spatially Fourier transform the resulting diffusion equation. The boundary condition at $z=d$ becomes homogeneous. The solution is given by
\begin{equation}
\label{eq:cs_full}
\Cs(r,z,t)-\Cl(t)=\frac{4}{\pi}\sum_{n=0}^\infty \frac{(-1)^n}{2n+1} f_n(r,t)\cos(\lambda_n z),
\end{equation}
with $\lambda_n=\frac{\pi}{d}\left(n+\frac{1}{2}\right)$ and 
\begin{equation}
\begin{split}
f_n(r,t)&= -\Cl(t)\\
\label{eq:f_n}
&+e^{-\Ds\lambda_n^2 t}\left[C_0(r)+\int_0^t \Cl(\tau)\Ds\lambda_n^2 e^{\Ds\lambda_n^2 \tau}\d{\tau}\right].
\end{split}
\end{equation}

Note that $C_0(r)$ is a constant if the film has not been patterned. This does not mean that in that case $f(r,t)$ does not depend on $r$, because we do not allow diffusion to take place for positions $r$ beyond the contact radius $a$ of the droplet, which depends on time. We presume that the compound diffusion ceases if the droplet is no longer present directly above the substrate, \textit{i.e.}, the concentration profile at a radial distance $r$ no longer evolves if the droplet's contact area radius $a(t)<r$. At any point in time, we know the value of $a$, as discussed above.

Equation~\eqref{eq:cs_full} describes how the compound concentration profile in the thin film evolves as a function of time due to the presence of a liquid droplet on top of the film. As a result of the simultaneous compound diffusion and droplet evaporation, the compound concentration in the liquid $\Cl$ changes as time progresses. We solve Eq.~\eqref{eq:cs_full} numerically as the droplet evaporates and find how the material inside the thin film is redistributed during the evaporation process. We return to this in Section~\ref{sec:watermarks}. 

If the concentration of the dissolved compounds exceeds the solubility limit, the surplus material precipitates, is deposited on top of the film and may pin the contact line, provided that the pinning force is non-zero. In the next section, we relate the deposit size and profile to the initial droplet properties. We show that the deposit size scales differently with the initial droplet size depending on the ratio between the timescales $\tauevap$ and $\taudiff$, associated with droplet evaporation and compound diffusion, respectively, and the purity of the initial liquid droplet. We find that the deposit profile decays linearly towards the centre of the deposit area. For the case where back-diffusion of material into the thin film is negligible, we obtain an analytical expression for the height of the deposit as a function of the radial distance.

\section{Deposit \textit{vs} droplet properties}
\label{sec:deposit}
Before presenting our results, we identify four different scenarios regarding the number of different compounds, the purity of the drop at time zero and the initial concentrations of the compounds within the thin film.
\begin{enumerate}
\item One species present in the thin film, and a pure liquid deposited onto the thin film;
\item One species present in the thin film, and the liquid containing the same species deposited onto the thin film;
\item One species present in the liquid deposited onto the thin film, and none in the film;
\item Multiple species present in the thin film and the liquid.
\end{enumerate}
Note that if we consider more than one component that this complicates matters considerably, for all of them might have different diffusivities, saturation concentrations in the liquid, give rise to different pinning behaviours, \textit{et cetera}. We for now consider a homogeneous distribution of compounds at time zero, $C_0(r)=C_0$, and set the equilibrium contact angle to $\thetaeq = \pi/2$, which is a typical value for water on a polymer film. We note that although we focus our attention on the droplet and the precipitate on top of the surface in this Section, we do in fact consider the exchange of soluble material with and the redistribution of material inside the thin film in our calculations. We postpone our in-depth discussions of both the time evolution of the material concentration in the film and initially patterned substrates, where $C_0(r)$ depends on the radial position, to Section~\ref{sec:watermarks}.

\begin{figure}[htp!]
\centering
\includegraphics[width=\figwidth]{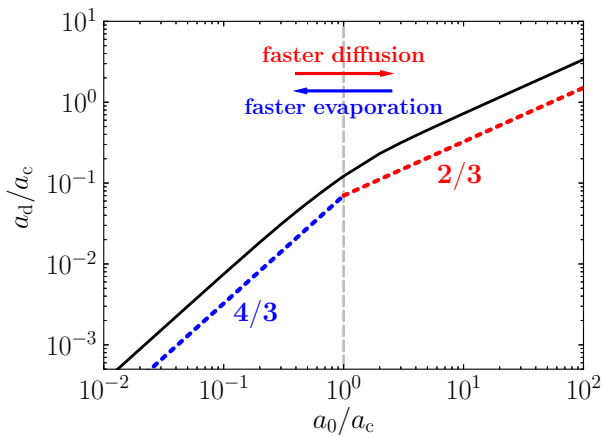}
\caption{The deposit size $\ad$ as a function of the initial droplet size $a_0$, both scaled to the characteristic size $\ac$, which is set by evaporation and diffusion. Two regimes are found, corresponding to fast evaporation $(a_0<\ac)$ and fast diffusion $(a_0>\ac)$. $C_0/\Csat = 0.5$.}
\label{fig:a0_vs_ad}
\end{figure}

In scenario 1, the thin film leaks a compound into the droplet that subsequently evaporates and concentrates that compound. If the concentration in the drop exceeds that of the film, diffusion back into the film takes place, redistributing the compound within the film. This aspect we discuss in Section~\ref{sec:watermarks} in more detail. If the concentration of compounds in the liquid exceeds its saturation value, the surplus material precipitates near the contact line, and the droplet becomes pinned temporarily or permanently, depending on the strength of the pinning force. This then defines the deposit radius $\ad$ of the debris left behind on the surface, see Fig.~\ref{fig:sketch_droplet}. In Fig.~\ref{fig:a0_vs_ad}, we have plotted our findings on the deposit size $\ad$ as a function of the initial droplet size $a_0$. Both radii are scaled to a critical radius $\ac$, defined as the size at which the characteristic timescales for evaporation and diffusion, $\tauevap$ and $\taudiff$, are equal,
\begin{equation}
\label{eq:a_c}
\ac = \sqrt{\frac{\Delta c}{\rho}\frac{\Dg}{\Ds}d^2},
\end{equation} see Eqs.~\eqref{eq:tau_evap} and \eqref{eq:tau_diff}. The compound diffusivity is presumably small, where $\Ds$ may reach values of order $10^{-15}\SI{}{\metre \squared \per \second}$ or smaller~\cite{Wei2009}. For typical values of the model parameters ($\Delta c\simeq 10^{-3}\SI{}{\kilo \gram \per \metre \cubed}$, $\rho\simeq 10^{3}\SI{}{\kilo \gram \per \metre \cubed}$, $\Dg\simeq 10^{-5}\SI{}{\metre \squared \per \second}$, $d\simeq 10^{-7}\SI{}{\metre}$), this results in a critical radius $\ac$ of order $1-\SI{10}{\micro \metre}$, while the droplet sizes relevant to lithography are in the order of tens to hundreds of micrometers~\cite{Chang2006, Kusumoto2006, Streefkerk2006}.

In Fig.~\ref{fig:a0_vs_ad}, we arbitrarily set $C_0/\Csat = 0.5$. Larger or smaller values lead to a vertical shift upward or downward of the shown curve. We note that the strength of the pinning force $\fp$ does not have an impact on the graph, since the deposit size $\ad$ is defined as the droplet size at the time that precipitation starts. The figure highlights two power law scaling regimes, representing fast diffusion and fast evaporation. For fast evaporation, $\ad \propto a_0^{4/3}$, and for fast diffusion we have $\ad \propto a_0^{2/3}$. We conclude that the deposit size depends differently on the initial droplet size, depending on whether diffusion or evaporation dominates the physics of the problem. In addition, the way that the deposit size depends on the film thickness $d$ is different for these two regimes: for fast diffusion, the deposit size increases with increasing film thickness, whereas for fast evaporation it decreases with increasing film thickness. This dependence of the deposit size on the film thickness is somewhat counter-intuitive but stems from the fact that the flux of material from the film into the support of the film is not allowed~\footnote{If we do not allow any flux of material at the support-film interface, it implies that the concentration gradient must be zero there. This boundary condition directly creates a non-locality of the solution for the diffusion equation, which results in the material flux at the film-liquid interface to be inversely proportional to the film thickness $d$.}. As a result, the flux of material from the film into the droplet is inversely proportional to $d$.

We recover the power law scalings described above by considering a simple scaling analysis. In the case of fast evaporation, we presume the flux density $J_z$ of material from the film into the droplet to be constant, $J_z = J_0$, since the concentration profile within the thin film has hardly any opportunity to evolve during the droplet evaporation. The amount of material $N$, dissolved in a hemispherical droplet at the time $\td$ at which the precipitation starts, can be calculated as
\begin{equation}
N = \int_0^{\td} \int_0^{a(t)} 2 \pi r J_0\, \d{r}\, \d{t} \simeq \frac{\pi \rho J_0}{4 \Dg \Delta c} a_0^4,
\end{equation}
for $\ad / a_0 \ll 1$, and follows directly from Eqs.~(\ref{eq:a_t}, \ref{eq:Lambda}). At $t=\td$, the concentration within the droplet has reached the saturation value, $\Cl = \Csat$, which is given by the ratio between the amount of dissolved material $N$ and the droplet volume ${\Vd = \frac{2}{3}\pi \ad^3}$,
\begin{equation}
\Csat = \frac{N}{\Vd} = \frac{3 \rho J_0}{8 \Dg \Delta c} \frac{a_0^4}{\ad^3}.
\end{equation}
Since $\Csat$ is a constant model parameter, we find
\begin{equation}
\ad = \left(\frac{3 \rho J_0}{8 \Dg \Delta c\, \Csat}\right)^{1/3} a_0^{4/3},
\end{equation}
for the case of fast evaporation. We note that since $J_0$ is inversely proportional to $d$, $\ad$ decreases with increasing film thickness as $d^{-1/3}$. 

In the case of fast diffusion, we presume that the film underneath the droplet becomes virtually fully depleted of soluble material, such that
\begin{equation}
N = C_0 d \pi a_0^2.
\end{equation}
This results in an expression for the deposit size $\ad$, given by
\begin{equation}
\ad = \left(\frac{3 C_0}{2 \Csat} d \right)^{1/3} a_0^{2/3}.
\end{equation}
From this scaling analysis, we recover the power law scalings for the deposit size that we find in Fig.~\ref{fig:a0_vs_ad}. Here, we only examined the dependence of the size of the deposit area $\ad$ on our model parameters. Next, we investigate in more detail how the deposited material is distributed onto the film after the precipitation is initiated.

\begin{figure}[htp!]
\centering
\includegraphics[width = \figwidth]{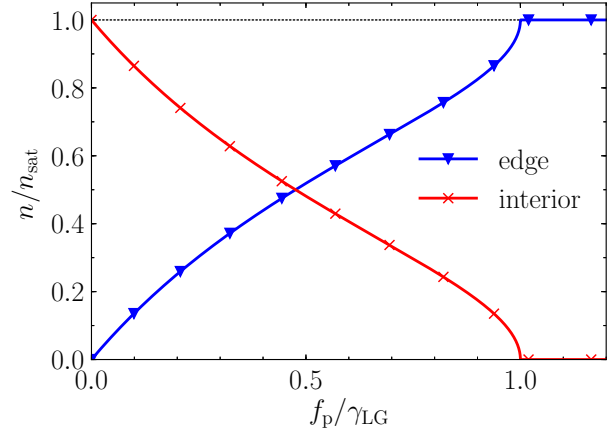}
\caption{Graphs for the amounts of material deposited near the edge (blue triangles) or in the interior (red crosses) of the deposit, relative to the total amount of material present in the droplet at $t=\td$, as a function of the dimensionless contact line pinning strength $\fp/\gammaLG$.}
\label{fig:deposit_edge}
\end{figure}

\begin{figure}[htp!]
\centering
\includegraphics[width = \figwidth]{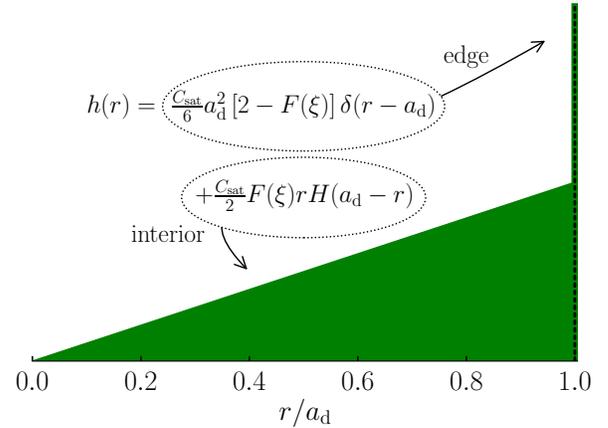}
\caption{Schematic of the topology of the deposit stain described by the amount per unit area $h(r)$ as a function of the scaled radial position $r/\ad$; $\xi \equiv \fp / \gammaLG$ is a dimensionless measure for the strength of the contact line pinning. See also the main text.}
\label{fig:deposit_topology}
\end{figure}

Within our model description, as soon as the precipitation occurs, the precipitate ends up near the contact line. The contact line may become pinned onto the deposit with a pinning force $\fp$. The strength of the pinning determines the deposit profile on top of the film; the stronger the pinning, the more material ends up near the edge of the deposit stain. From the point that the precipitation starts, the compound concentration in the droplet remains constant at $\Cl = \Csat$. If we neglect the back-diffusion of material from the droplet into the thin film, the amount of material $\nd$ that falls out of solution at a given time obeys
\begin{equation}
\label{eq:dndt}
\dd{\nd}{t} = - \Csat \dd{V}{t}.
\end{equation}
As long as the contact line remains pinned, all of the deposited material ends up near the edge of the deposit area. We can express the amount of material deposited near the edge $\nedge$ as
\begin{equation}
\label{eq:n_edge}
\nedge = -\int \Csat \d{V} = \Csat (\Vd - \Vdepin),
\end{equation}
where $\Vdepin$ denotes the volume of the droplet at the point of depinning. Equation~\eqref{eq:n_edge} represents the amount of dissolved material in the volume that is lost between the start of the deposition, where the droplet is hemispherical, and the depinning of the droplet. $\Vdepin$ is given by the volume of a droplet with a base radius of $\ad$ and a contact angle at its receding value $\theta = \thetar$. For $\thetaeq = \pi/2$ it reduces to,
\begin{equation}
\begin{split}
\Vdepin &= \pi \ad^3 \left(\frac{2-3\cos\thetar + \cos^3\thetar}{3\sin^3\thetar}\right) \\
&\equiv \tfrac{1}{3}\pi \ad^3\ F(\xi),
\end{split}
\end{equation}
where we have used Eq.~\eqref{eq:theta_r}; $\xi \equiv \fp / \gammaLG$ denotes the strength of the pinning with respect to the capillary forces, and
\begin{equation}
F(\xi) \equiv \frac{2 + \xi}{\left(1 + \xi \right)^{3/2}} \sqrt{1 - \xi}
\end{equation}
is a function that connects the dimensionless pinning force $\xi$ to the depinning volume $\Vdepin$. If $\xi = 0$ no pinning occurs, whereas if $\xi \geq 1$ the contact line becomes pinned permanently. The amount of dissolved material $\nedge$ that is deposited near the edge of the droplet then obeys
\begin{equation}
\label{eq:n_edge_final}
\frac{\nedge}{\nsat} = 1 - \frac{1}{2} F(\xi),
\end{equation}
with ${\nsat = \frac{2}{3}\pi \ad^3 \Csat}$ the amount of dissolved material present in the droplet at the time the deposition starts. In Fig.~\ref{fig:deposit_edge} we show the relative amounts of compound that are deposited near the edge (blue triangles) or in the interior (red crosses), as a function of the contact line pinning strength $\fp/\gammaLG$.
We learn from Eq.~\eqref{eq:n_edge_final} and Fig.~\ref{fig:deposit_edge} that the stronger the pinning force $\fp$, the more material ends up near the edge of the stain. The remainder of the dissolved material ends up in the interior of the deposit stain, after the contact line depins. As from this point on the contact angle $\theta$ remains constant at the receding value $\thetar$, the deposition profile of the interior is described by
\begin{equation}
\dd{n}{t} = -\Csat \dd{V}{a} \dd{a}{t},
\end{equation}
resulting in the amount of precipitate per unit area $\hint(r)$ decreasing linearly towards the centre of the stain,
\begin{equation}
\label{eq:h_int}
\hint(r) = \frac{1}{2 \pi r}\dd{n}{r} = \frac{\Csat}{2}F(\xi) r.
\end{equation}
We combine Eqs.~\eqref{eq:n_edge_final} and \eqref{eq:h_int} into an expression for the full deposit stain profile $h(r)$,
\begin{equation}
\label{eq:h_a}
h(r) = \frac{\Csat}{6}\ad^2\left[2-F(\xi)\right]\delta(r-\ad) + \frac{\Csat}{2} F(\xi) r H(\ad - r),
\end{equation}
see Fig.~\ref{fig:deposit_topology}. Here, $\delta(x)$ denotes the Dirac delta function, and the integration convention we use is ${\int_0^{\ad} \delta(r-\ad)\d{r} = 1}$. $H(x)$ denotes the Heaviside step function, where $H(x) = 1$ for $x > 0$ and $H(x) = 0$ for $x \leq 0$. 
Our findings on the topology of the deposit stain, with a pronounced edge and a decrease towards the centre, are qualitatively similar to experimental findings of~\citet{Belmiloud2012} on the topology of a watermark on a hydrophobic Si wafer. Note that the expressions presented above are valid as long as the back-diffusion of material from the droplet into the thin film is neglected. 

\begin{figure}[htp!]
\centering
\includegraphics[width = \figwidth]{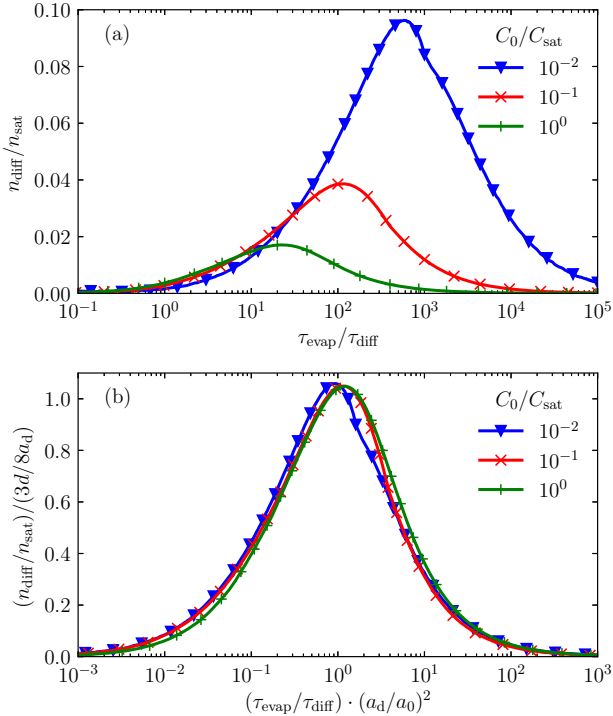}
\caption{The fraction of material that diffuses into the thin film after the saturation concentration is reached as a function of the relative diffusion rate. (a) $\ndiff / \nsat$ as a function of $\ratio$. (b) Both axes are rescaled appropriately with the deposit size $\ad$, resulting in a master curve for the diffused material as a function of the diffusion rate.}
\label{fig:ndiff_vs_ratio}
\end{figure}

If back-diffusion does occur, the amount of deposited material is smaller. We calculate the amounts of deposited and back-diffused material numerically and note that despite Eqs.~\eqref{eq:dndt}-\eqref{eq:h_a} are not strictly valid in this case, the general shape of the deposit profile remains the same. In fact, if the compound concentration inside the liquid droplet increases towards the solubility limit, the material simultaneously precipitates onto and diffuses back into the thin film. The fraction of material that diffuses back into the film depends on the relative diffusion rate $\ratio$, as well as on the initial compound concentration in the thin film $C_0/\Csat$. In Fig.~\ref{fig:ndiff_vs_ratio}a, we show the fraction of the amount of material that diffuses back into the film after the precipitation has started, $\ndiff$, and the amount of material present in the droplet when the precipitation starts, $\nsat$, as a function of the effective diffusion rate $\ratio$, for various initial concentrations $C_0/\Csat$, presuming that $\xi \geq 1$. In that case, the droplet becomes permanently pinned after the saturation concentration is reached.

We find that the fraction of material that diffuses back into the film during the precipitation exhibits a maximum at a certain value of $\ratio$. We can readily explain this by considering two extreme cases. Naturally, for slow diffusion ($\ratio \ll 1$), virtually all material is deposited on top of the film and no material diffuses into it. However, for fast diffusion ($\ratio \gg 1$), the dissolved material in the droplet and that in the thin film are in steady equilibrium. At the point where precipitation occurs, the compound concentration in the film is equal to that in the droplet and hence no net diffusion occurs. In between those two limits, material can diffuse back into the thin film. Since the initial compound concentration in the thin film $C_0$ strongly affects the size of the deposit $\ad$, $\ndiff$ is strongly influenced by the value of the former. In essence, the competition between the droplet evaporation and the compound diffusion after the saturation concentration is reached is now dictated by the diffusion time scale $\taudiff$ from Eq.~\eqref{eq:tau_diff}, and the typical time it takes for the saturated droplet to evaporate,
\begin{equation}
\left.\frac{\tauevap}{\taudiff}\right|_\mathrm{sat} = \frac{1}{\taudiff}\frac{\rho \ad^2}{2 \Dg \Delta c} = \frac{\tauevap}{\taudiff} \left(\frac{\ad}{a_0}\right)^2.
\end{equation}
The amount of material diffusing into the thin film after the saturation level is reached, that is, for $\left.\ratio\right|_\mathrm{sat} \approx 1$, may be estimated as
\begin{equation}
\frac{\ndiff}{\nsat} = \frac{J_z \pi \ad^2 \taudiff}{\frac{2}{3} \pi \ad^3 \Csat} \simeq \frac{\Ds \frac{\Csat}{2d} \pi \ad^2 \taudiff}{\frac{2}{3} \pi \ad^3 \Csat} = \frac{3d}{8\ad}.
\end{equation}
If we rescale the axes of Fig.~\ref{fig:ndiff_vs_ratio}a with the quantities above, we find that indeed the curves for different $C_0/\Csat$ collapse reasonably well onto a single master curve, see Fig.~\ref{fig:ndiff_vs_ratio}b. This indicates that the back-diffusion of the dissolved compounds is quasi-universal. However, both the amount and the timescales at which the process takes place are strongly affected by the saturation concentration. We note that in order to perform the rescaling in Fig.~\ref{fig:ndiff_vs_ratio}b, we need the deposit size $\ad$, which is a result from our calculations. Since the deposit size $\ad$ depends strongly and non-linearly on the ratios $\ratio$ and $C_0/\Csat$, a proper \textit{ab initio} rescaling of the axes cannot be found. In addition, we note that for the values for the ratio $C_0/\Csat$ depicted in Fig.~\ref{fig:ndiff_vs_ratio}, the evaporative and diffusive processes are in competition and are not directly affected by the finite thickness of the film. However, for smaller values of $C_0/\Csat$, virtually all soluble material in the film is taken up by the droplet. Conversely, if $C_0/\Csat > 1$, the material transport may not even be inverted for low to intermediate values of the compound diffusivity, as the compound concentration in the film remains higher than that in the droplet. In both limits, the universal behaviour of the amount of diffused material as a function of the competition between evaporation and diffusion breaks down.

\begin{figure}[htp!]
\centering
\includegraphics[width=\figwidth]{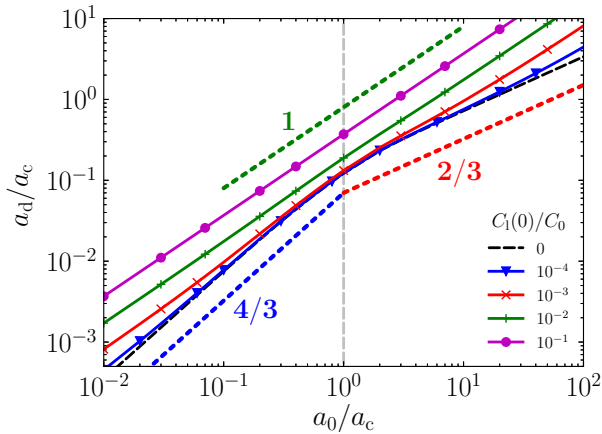}
\caption{The deposit size $\ad$ as a function of the initial droplet size $a_0$, both scaled to the characteristic size $\ac$, for various initial compound concentrations within the liquid drop $\Cl(0)/C_0$. A third regime with an exponent of unity arises, for high initial concentrations.}
\label{fig:a0_vs_ad_impure}
\end{figure}

This concludes our discussion of scenario 1, in which one species of compound is present in the thin film, and pure liquid is deposited onto the film. We now discuss to what extent the phenomena discussed above are of relevance to the alternative scenarios and in what aspects they exhibit different behaviours. Firstly, in the case of scenario 2, where initially both the film and the liquid contain the same species, the relative initial concentrations have a strong impact on the deposit size as a function of the initial droplet size. Due to the initial droplet already containing dissolved material, a third power law scaling arises. If we neglect the diffusion of material between the thin film and the droplet, the total amount of dissolved material $N$ contained in the droplet is,
\begin{equation}
N = \frac{2}{3}\pi a_0^3 \Cl(0),
\end{equation}
resulting in,
\begin{equation}
\ad = \left(\frac{\Cl(0)}{\Csat}\right)^{1/3} a_0.
\end{equation}
The introduction of this third regime affects the scaling behaviour that we showed in Fig.~\ref{fig:a0_vs_ad}. Depending on the initial compound concentration in the droplet, the size of the deposited droplet and the thickness of the film, we also find this regime. In Fig.~\ref{fig:a0_vs_ad_impure}, we show the deposit size $\ad$ as a function of the initial droplet size $a_0$, for various initial compound concentrations in the droplet $\Cl(0)/C_0$. For this plot we set again $C_0/\Csat = 0.5$.

Figure~\ref{fig:a0_vs_ad_impure} shows that the relation between the deposit size $\ad$ and the initial droplet size $a_0$ gradually transitions from two power law regimes with exponents $4/3$ and $2/3$ into a single power law regime with exponent $1$, as the initial compound concentration $\Cl(0)$ increases. Since the dimensions of the liquid droplet are much larger than the thickness of the thin film, $\Cl(0)$ only needs to be a fraction of the initial concentration in the thin film $C_0$ in order to have a considerable effect on the deposit size; the total dissolved amount in the initial droplet quickly becomes comparable or dominant to the amount present in the thin film.

Apart from the impact on the deposit size, the non-zero initial compound concentration in the drop of scenario 2 has hardly any effect on the properties of the deposit. The deposit topology depends largely on the strength of the contact line pinning (which we presume to be independent of the compound concentrations), and the diffusion of material from and into the thin film, governed by the characteristic timescales corresponding to the evaporation and diffusion processes. The same holds for scenario 3, in which the thin film is initially devoid of soluble compounds. In fact, in this case the relation between the deposit size and the initial droplet size yields a power law with exponent 1, regardless of the initial compound concentration in the drop. As no compounds are present in the thin film initially, all precipitating material originates from the initial droplet.

As we have hinted at before, in scenario 4, where multiple species of compound are present in both the thin film and the droplet, the deposition dynamics can become infinitely more complex. Generally, these compounds have different diffusivities, saturation levels or give rise to different pinning behaviours. If we presume that the diffusion processes of different compounds occur independently of each other, the precipitation of a single compound and the resulting contact line pinning may still affect the droplet shape. The deposition dynamics therefore becomes an interplay between all involved compounds. The contact line may even become pinned multiple times, with different pinning strengths, onto the different components that fall out of the solution. This is in addition to the mechanisms for multiple pinning events reported in the literature~\cite{Shmuylovich2002, Moffat2009, Kaya2010, Roy2015, Li2016}, which mainly consist of multiple deposition events of a single compound or colloidal particle type. The resulting footprint and its topology will in our case consist of several edges of the different deposited materials, effectively being a superposition of several profiles as presented in Fig.~\ref{fig:deposit_topology}. The amount of each compound diffusing between the droplet and the thin film is also directly affected, due to the size of the droplet's contact area being influenced by the multitude of components present in the droplet and the film. To discuss in detail the multitude of possible scenarios that arise with multiple components is beyond the scope of this work. We note that it is straightforward to incorporate these effects in our model. 

Until now, we have focused our attention on the size of the deposit area and the topology of the deposit stain. We discuss in the next section in more detail how the interior of the thin film is affected by the compound exchange between the film and the liquid droplet. 

\section{Compound redistribution inside the thin polymer film}
\label{sec:watermarks}
In the previous section we hinted at the fact that the fraction of material that is transported \textit{into} the thin film during the deposition \textit{onto} the film depends on the balance between diffusion and evaporation. However, as we shall see, the distribution of compounds within the thin film outside the deposition area also depends strongly on the diffusion process. We first consider an initially homogeneous thin film and show typical evolutions of the concentration profile within the film as a function of time. To highlight the effect the compound redistribution may have for, \textit{e.g.}, industrial applications, we subsequently examine a patterned substrate, where the compound concentration is a function of the radial position in the film. We consider an initially pure droplet, hemispherical in equilibrium, which at the point of saturation becomes permanently pinned onto the precipitated material that is deposited near the rim of the droplet.

\begin{figure}[htp!]
\centering
\includegraphics[width=\figwidth]{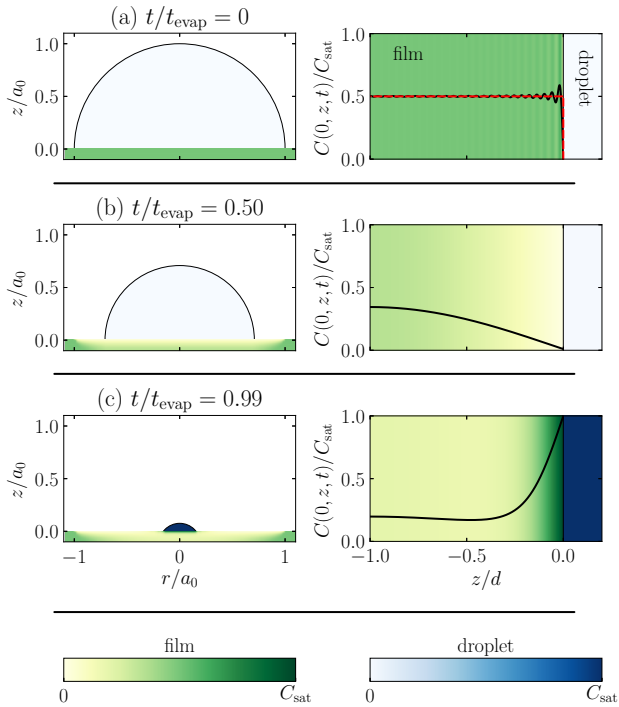}
\caption{The compound concentration profile inside the thin film $C(r,z,t)$ as a function of the radial and vertical positions $r$ and $z$, and the droplet at various points during the evaporation process: (a) $t/\tevap = 0$, (b) $t/\tevap = 0.50$, (c) $t/\tevap = 0.99$. $C_0/\Csat = 0.5$ and $\ratio=1$. $z$ and $r$ are both scaled to the initial droplet size $a_0$. Left: the compound concentration inside the droplet and the thin film. A darker colour represents a higher compound concentration in both the droplet and the film; Right: the compound concentration inside the thin film at the centre ($r=0$) of the droplet's contact area, as a function of the vertical position in the film $z/d$.}
\label{fig:conc_profile_homog}
\end{figure}

In Fig.~\ref{fig:conc_profile_homog}, we show the time evolution of the evaporating droplet and the concentration profile inside the thin film. We (again arbitrarily) set $C_0/\Csat = 0.5$ and $\ratio=1$, and note that in the Figure, the precipitated material is \textit{not} shown and that the thickness of the thin film is exaggerated for visualisation purposes, so is not to scale. Furthermore, for this Figure, note that we shifted the thin film down by a distance $d$, such that $z=0$ now corresponds to the top of the film and $z=-d$ to the bottom.
In Fig.~\ref{fig:conc_profile_homog}a we show the initial situation with a homogeneous compound distribution in the thin film and an empty droplet. We evaluate the concentration profile, described by Eq.~\ref{eq:cs_full}, taking into account the Fourier components in the sum up to $n=50$. The initial compound concentration is constant (depicted by the dashed red line), resulting in the strong Gibbs oscillations near the solid-liquid interface, if decomposed in 51 Fourier components. These do, however, die out quickly as the diffusion process starts, to which we return below. As the droplet is left to evaporate on the thin film, it takes up components from the film, see Fig.~\ref{fig:conc_profile_homog}b; the film slowly becomes depleted of compounds. The extent of the depletion depends on the compound diffusivity and the evaporation timescale and may be expressed in a diffusion depth,
\begin{equation}
\ddiff = \sqrt{2 \Ds \tauevap} = \sqrt{\frac{\rho}{\Delta c}\frac{\Ds}{\Dg} a_0^2},
\end{equation}
which is a measure for the depth into the thin film that is affected by the material transport between the film and the droplet. If $\ddiff = d$, the diffusion front reaches the thickness of the film and we transition from the ``fast evaporation'' to the ``fast diffusion'' regime of Fig.~\ref{fig:a0_vs_ad}; in fact, equating $\ddiff$ to $d$ is equivalent to equating the initial droplet size $a_0$ to the critical size $\ac$. Due to the difference between the droplet size and the film thickness, the compound concentration inside the droplet remains low until very late in the evaporation process. This has as a result that the material flux out of the film into the droplet decreases over time predominantly due to the limited amount of soluble material present inside the film, not due to the concentration increase inside the droplet. Later in the evaporation process, however, the concentration inside the droplet rises strongly due to the decrease of droplet volume, until it reaches the saturation value $\Csat$. At the point where the concentration inside the droplet exceeds the concentration just inside the film, the material flow inverts and the compounds are transported back into the film, see Fig.~\ref{fig:conc_profile_homog}c. We return to this phenomenon below. This final stage of the evaporation process results in a patch of concentrated compound inside the thin film around the centre of the droplet, near the solid-liquid interface.

\begin{figure}[htp!]
\centering
\includegraphics[width=\figwidth]{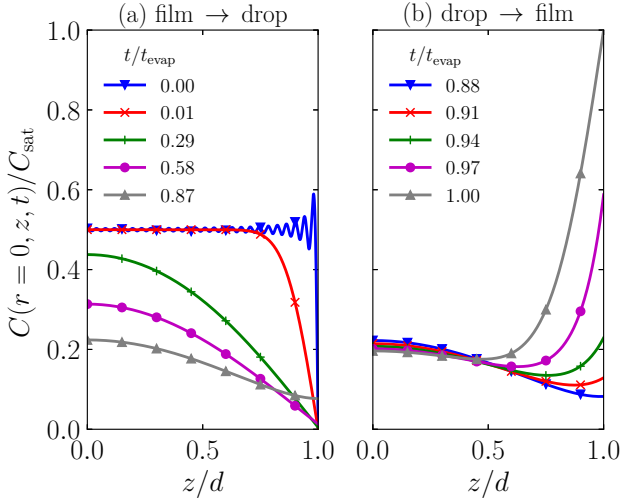}
\caption{The scaled compound concentration inside the thin film at the centre of the droplet's contact area ${C(r=0,z,t)/\Csat}$ as a function of the vertical position in the film $z/d$, for various points in time ${t/\tevap}$. ${C_0/\Csat = 0.5}$ and ${\ratio=1}$. (a) Net material transport occurs from the film into the droplet; (b) Net material transport occurs from the droplet into the film: back-diffusion.}
\label{fig:conc_profile_centre}
\end{figure}

As we evaluate the time evolution of the concentration profile more closely, we find that we can roughly divide it into two phases: (1) the compounds diffuse from the thin film into the droplet, and (2) the compounds diffuse back from the droplet into the thin film. In Fig.~\ref{fig:conc_profile_centre} we show the scaled concentration profile at the centre of the droplet's contact area $C(0,z,t)/\Csat$ as a function of the position in the film $z/d$, at various fractions of the total evaporation time of a droplet $\tevap$. We show the two phases (1) and (2) separately in Figs.~\ref{fig:conc_profile_centre}a and b. We note that the transition in time between the two phases depends on the value of $\ratio$, and remind the reader that at the film-droplet interface (located at $z/d = 1$) the compound concentration $C(r=0,z=d,t)$ is equal to that of the droplet $\Cl(t)$.

Figure~\ref{fig:conc_profile_centre}a confirms that the strong Gibbs fluctuations of the initial profile near the interface, and which constitute a modelling artefact, vanish completely within one percent of the evaporation time. If we include as many as 51 terms in the Fourier expansion of Eq.~\eqref{eq:cs_full}, we find a good approximation for the initial homogeneous distribution in the film. However, since the coefficients decay exponentially over time as $\Ds \lambda_n^2 t$, the higher orders become irrelevant virtually instantly. In fact, if we use only two Fourier orders to describe the process, much of the initial dynamics is still retained and the two solutions converge within 30\% of the droplet evaporation time. We note, however, that taking only so few orders into account may result in inaccuracies at late times, when the compound concentration in the droplet rises quickly and the profile develops strong gradients. We find that the concentration profiles flatten out gradually, as material is transported from the film to the droplet, while the compound concentration in the drop remains negligible. Only when $\Cl$ increases considerably, we find that the material flow is inverted and a transition is made into the second phase, shown in Fig.~\ref{fig:conc_profile_centre}b. A fraction of the dissolved compounds is transported back into the droplet, most of which ends up close to the film-droplet interface. Deep inside the film, however, we find hardly any effect from the back-diffusion. The concentration at the bottom of the film ($z / d = 0$) actually continues to decrease in the second phase, since it lags behind the temporal evolution of the concentration at the interface. The velocity of the ``concentration front'', shown in the right-hand side of Fig.~\ref{fig:conc_profile_centre}b, from the interface into the thin film, is again dictated by the compound diffusivity inside the film and hence depends on $\ratio$.

\begin{figure}[htp!]
\centering
\includegraphics[width=\figwidth]{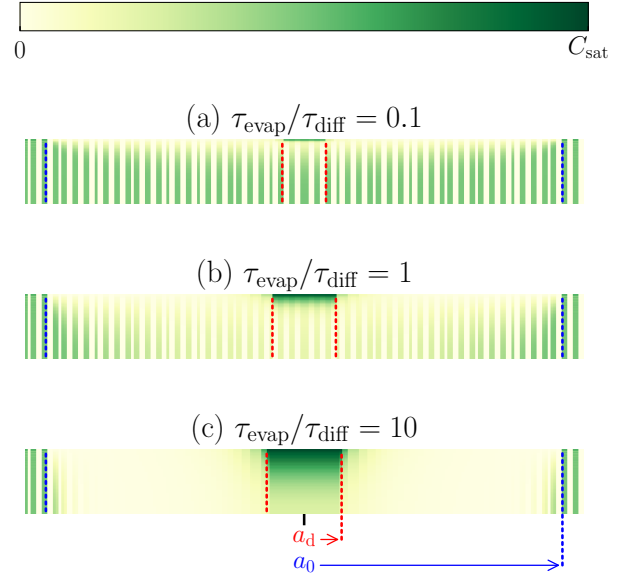}
\caption{The concentration profile inside the initially patterned thin film after the droplet has fully evaporated, for various values of $\ratio$: (a)~$0.1$, (b)~$1$, (c)~$10$. $C_0/\Csat = 0.5$. The initial droplet size $a_0$ (blue) and the deposit size $\ad$ (red) are indicated.}
\label{fig:conc_patterned}
\end{figure}

We have learned that an evaporating droplet on top of a thin film may redistribute the initially uniformly distributed material inside the film due to diffusion from the film into the droplet and back. The layer becomes locally partially depleted of compounds, which are redeposited onto and into the thin film in a concentrated patch around the centre of the droplet. In order to highlight and illustrate in more detail the effect this has on the final compound distribution, we consider a substrate that is initially patterned, \textit{i.e.}, the initial compound concentration $C_0(r)$ inside the film depends on the lateral position in the film. This is the case, \textit{e.g.}, after irradiation of the photoresist film in photolithography. Our patterning consists of regular, radially equally spaced lines that contain alternating high and low compound concentrations. We set the high concentration to a constant $C_0$, while choosing the low concentration to be zero. In Fig.~\ref{fig:conc_patterned} we show the influence of the value of $\ratio$ on the final distribution of compounds in the thin film, for $C_0/\Csat = 0.5$.
It clearly demonstrates the impact the value of $\ratio$ has on the final compound distribution after the droplet has completely evaporated. We can see in Fig.~\ref{fig:conc_patterned}a that for small values of $\ratio$, \textit{i.e.}, fast evaporation, the film hardly becomes depleted of compounds and the material redistribution remains (literally) superficial. As we increase the value of $\ratio$ in Figs.~\ref{fig:conc_patterned}b and c, the diffusion front penetrates deeper into the film, which eventually becomes almost completely depleted of compounds, as these diffuse into the droplet. To further increase the diffusivity of compounds inside the film would result in a similar picture due to the finite nature of the film: only a finite amount of material is present to be redistributed.

Despite the differences between the final concentration profiles for different values of $\ratio$, we can identify some similarities between them. In all three profiles, the contrast (\textit{i.e.}, the concentration difference between the high-concentration and low-concentration regions) of the line pattern decreases towards the centre of the stain, albeit to different extents. We note, however, that close to the film-droplet interface, the patterns are strikingly similar. 

\begin{figure}[htp!]
\centering
\includegraphics[width=\figwidth]{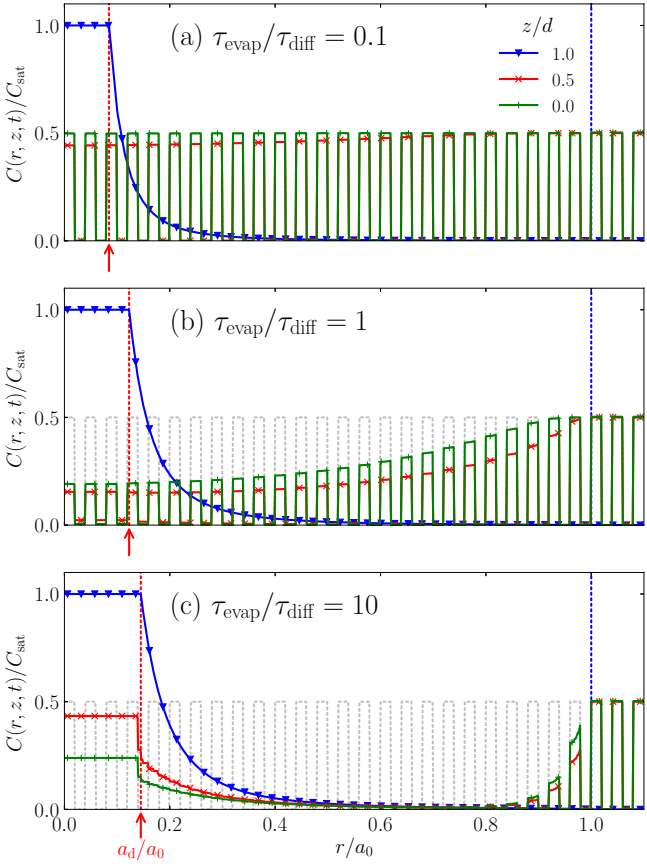}
\caption{The concentration profiles inside the initially patterned thin film after the droplet has fully evaporated, for various values of $\ratio$: (a)~$0.1$, (b)~$1$, (c)~$10$. $C_0/\Csat = 0.5$. A cross-section is taken at various heights in the film: at the top ($z/d = 1.0$, blue triangles), the middle ($z/d = 0.5$, red crosses) and the bottom ($z/d = 0.0$, green pluses). The compound concentration $C(r,z,t)/\Csat$ is shown as a function of the radial position $r/a_0$. The dashed lines indicate the unperturbed pattern (grey), the initial droplet size $a_0$ (blue) and the deposit size $\ad$ (red).}
\label{fig:conc_profile_radial_end}
\end{figure}

We investigate these phenomena in more detail in Fig.~\ref{fig:conc_profile_radial_end}, where we show the compound concentration $C(r,z,t)/\Csat$ as a function of the radial position $r/a_0$, for various values of $\ratio$ and at various heights in the thin film $z/d$.
In all three Figs.~\ref{fig:conc_profile_radial_end}a, b and c, we find that at the interface ($z/d = 1.0$, blue triangles), the compounds are taken up by the liquid droplet, after which they are redeposited into the film near the centre of the droplet's contact area. Toward the centre of the stain, we see an increase of compound concentration until it reaches the saturation value $\Csat$ inside the deposition area ($r \leq \ad$), effectively blocking the underlying patterning from reaching the surface. We can recognise this phenomenon in Fig.~\ref{fig:conc_patterned} in the concentrated patch of material near the interface in the deposit area. If we look deeper into the film, however, for small values of $\ratio$ (Fig.~\ref{fig:conc_profile_radial_end}a), we find that in the middle of the film ($z/d = 0.5$) the pattern has only been affected slightly, while at the bottom ($z/d = 0.0$) it remains virtually unaffected by the droplet. Upon increasing the diffusivity relative to the evaporation rate, in the case that $\ratio = 1$ (shown in Fig.~\ref{fig:conc_profile_radial_end}b), we see that the concentration profile is clearly affected by the evaporating droplet, all the way down to the bottom of the film. Note, however, that although the contrast has decreased, the original patterning may be recovered from the film: already at $z/d = 0.5$ the regions of high and low concentration may still be distinguished. This changes for greater values of $\ratio$, as we show in Fig.~\ref{fig:conc_profile_radial_end}c. Due to the relatively fast diffusion, the film is fully depleted of the soluble compound. Towards the deposition area ($r \leq \ad$), the compound concentration in the thin film increases due to back-diffusion, however, all information on the initial patterning is lost. 

We conclude from this that for modest values of the ratio $\ratio$, the disruption of the concentration profile remains fairly superficial, and hence the original pattern may be recovered by removing the top part of the film. For greater values, however, the patterning is lost unrecoverably.

\section{Summary and Conclusions}
\label{sec:conclusions}
In conclusion, we put forward a macroscopic model for the redistribution of compounds from a thin film due to the presence of a sessile evaporating droplet. We identify four different scenarios, related to the number of compounds present and their initial concentrations in the film and in the droplet. We presume that vertical diffusion is the predominant mode of transportation in the material. The compounds present in the film diffuse into the liquid droplet on top of the substrate. In the late stages of the evaporation process, the concentration within the droplet strongly rises until the saturation level is reached. At this point, dissolved compounds start to fall out of solution and are deposited near the contact line of the drop, on which the contact line may become pinned, temporarily or permanently. The nature of the pinning depends on the strength of the pinning force of the contact line on the precipitate. In addition, the compounds diffuse back into the thin film, reversing the diffusive transport.

For the case that we consider only one species of compound that originates entirely from the thin film, we find two power law regimes for the size of the deposition area as a function of the initial droplet size. In the fast-evaporation limit, the deposit size scales with the initial droplet size to a power $4/3$, whereas for fast diffusion, this exponent is $2/3$. It turns out that the topology of the resulting deposit strongly depends on the strength of the contact line pinning onto the precipitate. If the pinning is stronger, more material ends up near the contact line and less of it is deposited in the interior of the deposit stain. This results in a strongly pronounced edge, and a linearly decreasing profile towards the centre of the stain. During the deposition of material, a fraction of the dissolved compounds diffuse back into the thin film. This amount exhibits a maximum, the magnitude and position of which are strongly dependent on the deposit size and therefore on the initial concentration of compounds in the film.

If some dissolved compounds are already present inside the droplet from the start, a third power law regime arises. Depending on the initial concentration in the drop, the deposit size becomes linearly proportional to the initial droplet size. Apart from the effect on the deposit size, an impure initial droplet does not affect the deposition process qualitatively. This also holds in the case of an empty initial film and an impure droplet, but in this case we only find an exponent of $1$ for the scaling between the initial droplet size and deposit size. If multiple compounds are involved, the deposition process becomes more complex, as all of the compounds generally exhibit different diffusivities, saturation concentrations and may result in different contact line pinning and depinning events. We do not discuss this in detail in this work. However, it can be captured by our model in a straightforward fashion.

We find that in addition to a deposition stain on top of the thin film, an evaporating droplet may affect the initially uniform compound distribution in the film. We find that in general, components diffuse from the thin film into the droplet, partially depleting the thin film from compounds. In the late stages of the evaporation, however, the concentration in the droplet increases strongly and a portion of the compounds diffuses back into the film, inverting the material transport. This effect becomes more pronounced if we consider patterned substrates, where the initial compound concentration depends on the lateral position in the film, such as is the case after the illumination of the photoresist film in photolithography. During the evaporation process, the contrast (\textit{i.e.}, the difference between the high-concentration and low-concentration regions) of the pattern decreases from the outside towards the centre and the patterning of the film becomes fully blocked from reaching the surface in the deposition area. This is in qualitative agreement with experimentally measured watermark defects in immersion lithography~\cite{Chang2006, Wei2009}. Depending on the compound diffusivity, this effect may or may not remain superficial. This implies that for modest values of the diffusivity, not all of the film is affected by the droplet and the original pattern may be recovered by removing a fraction of the film from the top.

\begin{acknowledgments}
This work is part of the research programme ``Towards zero defectivity'' with project number 13919, which is partly financed by the Netherlands Organisation for Scientific Research (NWO).
\end{acknowledgments}

\bibliography{bibliography}

\end{document}